\documentstyle[multicol,aps,prb,epsf]{revtex}

\begin{document}

\draft

\title{
Spectral function of the spiral spin state in the 
trestle and ladder Hubbard model 
}

\author{
Ryotaro Arita
}
\address{Department of Physics, University of Tokyo, Hongo,
Tokyo 113-0033, Japan}
\author{
Koichi Kusakabe
}
\address{Institute for Solid State Physics, University of Tokyo,
Roppongi, Tokyo 106-8666, Japan}
\author{
Kazuhiko Kuroki and Hideo Aoki
}
\address{Department of Physics, University of Tokyo, Hongo,
Tokyo 113-0033, Japan}

\date{\today}

\maketitle

\begin{abstract}
Eder and Ohta have found a violation of the Luttinger rule 
in the spectral function 
for the $t$-$t'$-$J$ model, which was interpreted as
a possible breakdown of the Tomonaga-Luttinger(TL) description 
in models where electrons can pass each other. 
Here we have computed the spin correlation along with the 
spectral function for the one-dimensional $t$-$t'$ Hubbard model 
and two-leg Hubbard ladder.  By varying the Hubbard $U$ 
we have identified that such a phenomenon is in fact 
a spinless-fermion-like behavior of holes moving in 
a {\it spiral} spin configuration that has 
a spin correlation length of the system size.
\end{abstract}

\medskip


\begin{multicols}{2}
\narrowtext

It is widely believed that the low-energy physics of 
a wide class of one-dimensional (1D) systems can be described as a 
Tomonaga-Luttinger(TL) liquid\cite{TL1,TL2}, 
which is an effective theory for electrons interacting 
in 1D. 
The ansatz has indeed been shown to be valid 
for exactly solvable 1D models such as the Hubbard model or 
the supersymmetric $t$-$J$ model, and also for some other models numerically. 
However,
recently, Eder and Ohta\cite{EderOhta} looked into the spectral function 
in 1D $t$-$J$ model that has $t'$ ($t$-$t'$-$J$ model), and 
made a very interesting observation that the density of electrons $n$ and 
Fermi momentum $k_F$ are related with $k_F=\pi n$ in a certain
parameter regime, which is incompatible with the Luttinger relation, 
$k_F=\pi n/2$, expected for a TL liquid, which they suggest to be 
an indication for a breakdown of the TL description. 

In order to clarify the origin of such an curious behavior,
here we study the $t$-$t'$ Hubbard model with finite values of $U$.  
The reason we have chosen the Hubbard model 
is that the magnetic phase diagram on the $n-U$ plane has been obtained 
for the $t$-$t'$ 
Hubbard model by Daul and Noack (inset of Fig.2),\cite{Daul,Daul2} 
so that we can identify the region on which we work.  
For the Hubbard model we find the same curious behavior of the spectral 
function.  We further find that the state in question 
is a {\it spiral} spin state, 
in which the spin correlation has a wave length of the system size
and thus has a local ferromagnetic nature.
The curious behavior of the spectral function
can be understood by a picture in which holes can 
hop almost freely in such a spin background as originally
proposed by Doucot and Wen\cite{DoucotWen}
as a trial state for finite systems
to prove the instability of Nagaoka's ferromagnetism\cite{Naga}.
The ferromagnetic-like state naturally explains the doubling of 
$k_F$ into $\pi n$ as a spinless-fermion-like behavior. 
The region over which the 
spiral behavior appears is indeed 
consistent with the phase diagram for the $t$-$t'$ 
Hubbard model. 
In order to see if the appearance of the spiral state extends to 
other quasi-1D systems, 
we have also studied the 2-leg Hubbard ladder with $U=\infty$,
and have found similar features as in the $t$-$t'$ Hubbard model.

The $t$-$t'$ Hubbard Hamiltonian is given by 
\begin{eqnarray*}
{\cal H}&=&-t\sum_{i=1}^{L}\sum_{\sigma}
(c_{i \sigma}^{\dagger}c_{i+1,\sigma}+{\rm H.c.})\\
&&+t'\sum_{i=1}^{L}\sum_{\sigma}
(c_{i \sigma}^{\dagger}c_{i+2,\sigma}+{\rm H.c.})
+U\sum_{n=1}^{L}n_{i \uparrow}n_{i \downarrow},
\end{eqnarray*}
in standard notations. Hereafter we set $t=1$. 

First we numerically calculate, 
with the continued fraction expansion, 
the single-particle spectral function given by
\begin{eqnarray}
A^{\pm}(k,\omega)=\frac{1}{\pi}{\rm Im}
\langle \Phi_G | \gamma^{\mp}_{k \sigma}
\frac{1}{\omega\pm(E_0-{\cal H})-i0}
\gamma^{\pm}_{k \sigma}|
\Phi_G \rangle,
\label{akwdef}
\end{eqnarray}
where $A^{+}(A^{-})$ denote the electron addition (removal) spectrum 
with $\gamma^{+}_{k \sigma}\equiv c^{\dagger}_{k \sigma}$, 
$\gamma^{-}_{k \sigma}\equiv (\gamma^{+}_{k \sigma})^{\dagger}$, 
$| \Phi_G \rangle$ and $E_0$ the ground state and its energy, respectively. 
In Fig. \ref{akw1}, we show the results for the case of
10 electrons on a 12 site ring, $U=40 (a)$ and $U=20 (b)$ 
for $t'=0.2$.
We can see that for $U=20$, $k_F=\pi n/2$ ($n=10/12$) is satisfied, while
the equality is violated in favor of $k_F=\pi n$ for $U=40$, 
which would be expected for spinless fermions. 
Such a behavior persists for larger values of $U$. 
Thus the situation is indeed similar to the result for the $t$-$t'$-$J$ model 
obtained by Eder and Ohta\cite{EderOhta}.

At this stage, let us recall the phase diagram of the $t$-$t'$ 
Hubbard model obtained by 
Daul and Noack\cite{Daul,Daul2} with 
the density matrix renormalization group (DMRG) method 
in open boundary conditions.
For $t'=0.2$, $n\sim 1$ and large $U$, they
found a large ferromagnetic region in the phase diagram.
If the periodic boundary condition(PBC) is assumed, on the other hand, 
the ground state is $S=0$ for any value of $U$ at least for system 
sizes up to 12 sites 
as also pointed out by Daul and Noack\cite{Daul2}.
We find here that the region in the $n-U$ plane 
over which the 
spinless-fermion-like behavior of $A(k,\omega)$ is observed 
roughly coincides with the ferromagnetic 
region in the Daul-Noack phase diagram (inset of Fig.2)\cite{apbccomm}.

Such an observation has motivated us to look into the spin-spin
correlation function, 
$\langle \Phi_G | S_i^z S_j^z | \Phi_G \rangle$, 
to identify the nature of the $S=0$ ground state in PBC.
We show in Fig. \ref{spinspinco} the result for 
10 electrons in a 12-site ring with $U=20,40$.
We can see that for $U=40$, which belongs to 
the spinless-fermion-like region, 
the spin-spin correlation indeed indicates the spiral spin state 
with the spin correlation being as large as the system size.
Such a spiral spin-spin correlation has also been 
encountered rather ubiquitously in
1D for Tasaki's model\cite{watamiya}, 
double exchange model\cite{Kubo}, 
the Kondo lattice model\cite{Sigrist}, and
the 2-band Hubbard model\cite{Kusakabe94}
and also in the 2D Hubbard model with $U=\infty$ for 
two holes\cite{Kusakabe}.

Nature of such a spiral state has been discussed 
by Doucot and Wen(DW)\cite{DoucotWen}, who studied
the infinite-$U$ Hubbard model with two holes, and 
found a trial state which gives an energy lower 
by inverse system size
than the Nagaoka's ferromagnetic state.  
The key idea of DW is based on the following intuition.  
Holes behave as free fermions for on-site interactions 
when the background spin state is ferromagnetic (or nearly so). 
Then we have only to worry about the fermion statistics, i.e., 
the antisymmetry (node) in the wave function against 
the exchange of two holes.
If we impose the node upon the spin part,
the kinetic energy of holes can be lowered.
This is accomplished by twisting the spin alignment in a full turn, 
but very slowly over the entire sample dimension 
to minimize the cost in energy and 
to maintain the ferromagnetic nature.
Another way of saying is that the spiral spin texture generates
a fictitious gauge flux which absorbs the frustration induced by
the fermion statistics of the holes.  
Indeed the flux mimics an Aharonov-Bohm of half flux quantum 
to shift the k-points by half the k-point spacing to let an even number of
spinless fermions take a closed shell configuration.

While these are conceived for the ordinary Hubbard model, 
it is intriguing to study whether 
the spinless-fermion-like behavior of $A(k,\omega)$ found here for 
the quasi-1D system 
can be understood quantitatively in terms of the DW state.
Although in the original paper DW considered
the states in which
the holes are dressed by spin waves as well, 
let us consider the state where holes hop in a rigid spin 
background for simplicity.  
The energy of the hole in the $t$-$t'$ Hubbard model
is then given as 
\begin{eqnarray}
\varepsilon&=&-2t\cos\left(\frac{\pi}{L}\right)
\cos\left(\frac{2N\pi}{L} \pm \frac{\pi}{L}\right) \nonumber\\
&&+2 t' \cos\left(\frac{2\pi}{L}\right)
\cos\left(\frac{4N\pi}{L} \pm \frac{2 \pi}{L}\right),
\label{energy}
\end{eqnarray}
where $N$ is an integer with $0<N<L$ and $\pm$ corresponds 
to the sign of the fictitious flux.  
In Fig. \ref{akw1}(a), we plot the energy dispersion defined by
the equation (\ref{energy}).
We can see that the single-particle spectrum is reproduced 
remarkably well.

In the result for the addition spectrum 
an almost dispersionless band
of low intensity peaks is seen (at around $E=2$ in Fig. \ref{akw1}(a)). 
If the ground state were a fully spin-polarized 
ferromagnetic state, then such a band is trivially expected, since 
we can add an opposite spin at any $k$-point.  
Remnant of such a band 
again suggests the ferromagnetic nature of the present ground state. 

A next question is whether the spiral state persists 
for more than two holes.  
In Fig. \ref{spinspinco2}, we show the spin-spin correlation
for 6 electrons on a 12-site lattice ($n=0.5$; quarter filled) in PBC 
for $U=4,6$.  
We can see that for $U=6$, the spin-spin correlation is again spiral.
To investigate whether Doucot and Wen's picture is 
valid for such an intermediate doping,
we have calculated the single-particle spectrum 
for $U=6$ in Fig. \ref{akw2}(a), 
$U=4$ in Fig. \ref{akw2}(b).
We can see that 
the nature of the spectrum also changes between $U=4$ and $U=6$.  
The spectrum for the spiral state is fitted well by the energy dispersion
defined by the eq.(\ref{energy})(dashed curves in Fig. \ref{akw2}(a)) 
even though the hole concentration is as large as $n_h=0.5$.
This is surprising, since the assumption that
holes are nearly free
would be valid only for infinitesimal doping.   
For detailed energetics the sample size dependence need be studied.

We next question 
whether the antiferromagnetic state is
connected adiabatically to the spiral state as we increase the Hubbard $U$.  
We show the ground-state energy as a function of $U$ for 
6 electrons on a 12-site ring in Fig. \ref{level}.
We can clearly identify a level crossing appearing as a cusp 
around $U \sim 5$, 
which indicates that a transition occurs within the $S=0$ 
space from the antiferromagnetic
phase (low-$U$ side) to the spiral phase (high-$U$ side).  

Finally, we move on to the 2-leg Hubbard ladder.  
Ferromagnetic behavior is also found in the ladders where
$U$ is infinite\cite{Liang,Kohno},
so it is intriguing to study whether a spiral 
state appears in this system as well, and if so, whether 
$A(k,\omega)$ exhibits a DW-like behavior.
The Hamiltonian is given by
\begin{eqnarray*}
{\cal H}&=&-t\sum_{i=1}^{L}\sum_{\alpha,\sigma}
(c_{i\alpha \sigma}^{\dagger}c_{i+1\alpha\sigma}+{\rm H.c.}) \\
&&-t\sum_{i=1}^{L}\sum_{\sigma}
(c_{i1\sigma}^{\dagger}c_{i2\sigma}+{\rm H.c.})
+U\sum_{n=1}^{L}\sum_{\alpha}n_{i \alpha \uparrow}n_{i \alpha \downarrow},
\end{eqnarray*}
where $\alpha(=1,2)$ labels the two legs of the ladder.
We set $t=1$, $U=\infty$ and we consider only the case of
two holes.
We have calculated the spin-spin correlation $\langle S_{i \alpha}^z
S_{j \alpha}^z \rangle$ (not shown), and found that the ground state is
indeed spiral. We have also calculated the single-particle excitation, 
eq.(\ref{akwdef}), where we now take 
$\gamma^{+}_{k\sigma}=c^{\dagger}_{k 1\sigma}+c^{\dagger}_{k 2\sigma}$
(bonding states) or 
$\gamma^{+}_{k\sigma}=c^{\dagger}_{k 1\sigma}
-c^{\dagger}_{k 2\sigma}$(antibonding states).
In Fig. \ref{akwlad}, we show the result for two holes (14 electrons on
a $8 \times 2$ ladder).
We fit the spectrum by a two-band extension of eq.(2),
\begin{eqnarray}
\varepsilon=-2t\cos\left(\frac{\pi}{L}\right)
\cos\left(\frac{2N\pi}{L} \pm \frac{\pi}{L}\right) \pm t.
\label{energy2}
\end{eqnarray}
Fig. \ref{akwlad} shows that the picture of two holes hopping 
in a twisted spin background is surprisingly accurate.

To summarize, we have pointed out that 
the behavior of the spectral function 
found by Eder and Ohta for the $t$-$t'$-$J$ model 
can be understood as a realization of the spiral spin state.   
As mentioned above, the fully polarized ferromagnetic state is 
the ground state for the open-boundary 
$t$-$t'$ Hubbard model\cite{Daul,Daul2}.
This is also the case with the open-boundary Hubbard ladders\cite{Liang,Kohno}.
Then, in the thermodynamic limit, 
the spiral ground state (in PBC) can either give way to the ferromagnetic 
state, remain the ground state, or become degenerate with the 
ferromagnetic one. 
This depends only on whether the cost to twist the spin configuration 
is finite or tends to zero with the sample size, since 
the energy gained by accommodating the electrons in 
a closed-shell configuration vanishes in the thermodynamic limit.
Hence an important future problem is to clarify the
relation between the spiral state and the ferromagnetic state 
in the thermodynamic limit and their spin stiffness.

We thank R. Eder and Y. Ohta for a critical reading of the manuscript.  
Numerical calculation were done on SR2201 at the
Computer Center of the University of Tokyo and
FACOM VPP 500/40 at the Supercomputer center, Institute
for Solid State Physics, University of Tokyo.

\begin{figure}
\caption{
Full single particle spectral function for 10 electrons on a 
12 site $t$-$t'$ lattice, U=40 (a) and U=20 (b).
Solid line denotes the electron removal spectrum, while 
dotted line the electron addition spectrum. 
Dashed curves in (a) indicate the energy dispersion defined by
eq.(2).
}
\label{akw1}
\end{figure}

\begin{figure}
\caption{
The spin-spin correlation function for 10 electrons on a 12 site lattice,
$U=20$ (solid line) and 40 (dashed line)
with the periodic boundary condition.
The inset shows the phase boundary of the
ferromagnetic phase (solid curve)
due to Daul and Noack$^{2,3}$
on which the present identification of the spiral phase (solid 
straight lines) 
is indicated for $n=10/12$ (Figs. 1,2) and $n=6/12$ (Figs. 3,4).
}
\label{spinspinco}
\end{figure}

\begin{figure}
\caption{
A plot similar to Fig. 2
for 6 electrons 
on a 12 site $t$-$t'$ lattice,
$U=4$ (solid line) and 6 (dashed line).
}
\label{spinspinco2}
\end{figure}

\begin{figure}
\caption{
A similar plot to Fig. 1
for 6 electrons on a
12 site $t$-$t'$ lattice, $U=6$(a) and $U=4$(b)
}
\label{akw2}
\end{figure}

\begin{figure}
\caption{
The ground state energy for 6 electrons on a 12 site $t$-$t'$ lattice
as a function of $1/U$.
}
\label{level}
\end{figure}

\begin{figure}
\caption{
A similar plot to Fig. 1 
for 14 electrons on a
$8 \times 2$ ladder.
Dashed curves indicate the dispersion defined by 
eq.(3).
}
\label{akwlad}
\end{figure}

\end{multicols}

\begin{references}
\bibitem{TL1}J. S\'{o}lyom, Adv. Phys. {\bf 28}, 201 (1979);
V.J. Emery, in {\it Highly Conducting One Dimensional Solids} 
ed. by J.T. Devreese {\it et al.} (Plenum, 1979), p. 247.
\bibitem{TL2}H. J. Schulz, in {\it Mesoscopic Quantum Physics},
ed. by E. Akkermans {\it et al.} (Elsevier, 1995), p. 533; 
J. Void, Rep. Prog. Phys. {\bf 57}, 977 (1994).
\bibitem{EderOhta}R. Eder and Y. Ohta, Phys. Rev. B {\bf 56}, R14247 (1997).
\bibitem{Daul}S. Daul and R. M. Noack, Z. Phys. B {\bf 103}, 293 (1997).
\bibitem{Daul2}S. Daul and R. M. Noack, Phys. Rev. B {\bf 58}, 2635 (1998).
\bibitem{apbccomm}As Daul and Noack$^5$ pointed out, if we take 
anti-periodic boundary condition, the critical $U$ for 
the ferromagnetic-nonmagnetic boundary is underestimated
for $n\rightarrow 0$.
For intermediate values of $n$, we can show that 
the ferromagnetic region for the anti-periodic boundary 
roughly coincides with that for the open boundary 
as will be published elsewhere.
\bibitem{DoucotWen}B. Doucot and X.G. Wen, Phys. Rev. B 
{\bf 40}, 2719 (1989).
\bibitem{Naga}Y. Nagaoka, Phys. Rev. {\bf 147}, 392(1966).
\bibitem{watamiya}Y. Watanabe and S. Miyashita,
J. Phys. Soc. Jpn. {\bf 66}, 3981 (1997).
\bibitem{Kubo}K. Kubo, J. Phys. Soc. Jpn. {\bf 51}, 782 (1982).
\bibitem{Sigrist}M. Sigrist, K. Ueda and H. Tsunetsugu,
Phys. Rev. Lett {\bf 67}, 2211 (1991).
\bibitem{Kusakabe94}K. Kusakabe and H. Aoki, Physica B
{\bf 194-196}, 217 (1994). 
\bibitem{Kusakabe}K. Kusakabe and H. Aoki, Phys. Rev. B {\bf 52},
R8684 (1995).
\bibitem{Liang}S. Liang and H. Pang, Europhys. Lett. {\bf 32}, 173 (1995).
\bibitem{Kohno}M. Kohno, Phys. Rev. B {\bf 56}, 15015 (1997).
\end{references}
\end{document}